\documentclass[aps,prb,a4paper,reprint,superscriptaddress,showpacs]{revtex4-1}
\usepackage{graphicx,graphics,color,epsfig}
\usepackage{epstopdf} %This line makes .eps figures into .pdf - please comment out if not required.
\usepackage{bm}
\usepackage{hyperref}
\usepackage{amsmath}
\usepackage{amsfonts}
\usepackage{amssymb}
\usepackage{appendix}
\usepackage{booktabs}

\begin{document}

\title{Electric-field-induced extremely large change in resistance in graphene ferromagnets}

\author{Yu Song}\thanks{Corresponding author}
%\email{songyu@mtrc.ac.cn}
\email{kwungyusung@gmail.com}
\affiliation{Microsystem and Terahertz Research Center, China Academy of Engineering Physics,
Chengdu 610200, P.R. China}
\affiliation{Institute of Electronic Engineering, China Academy of Engineering Physics,
Mianyang 621999, P.R. China}

\begin{abstract}
A colossal magnetoresistance ($\sim 100\times10^3\%$) and an extremely large magnetoresistance ($\sim 1\times10^6\%$) {have been previously} explored in manganite perovskites and Dirac materials, respectively. However, the requirement of an extremely strong magnetic field (and {an} extremely low temperature) makes them not applicable for realistic devices. In this work, we propose a device that can generate even larger changes in resistance in a zero-magnetic field and at {a} high temperature. The device is composed of a graphene under two strips of yttrium iron garnet {(YIG)}, where {two} gate voltages are applied to cancel the heavy charge doping in the {YIG-}induced half-metallic ferromagnets. {By calculations using the Landauer-B\"{u}ttiker formalism, we} demonstrate that, when a proper gate voltage is applied on the free ferromagnet, changes in resistance up to $305\times10^6\%$ ($16\times10^3\%$) can be {achieved} at the liquid helium (nitrogen) temperature and in a zero magnetic field. We attribute such a remarkable effect to a gate-induced full-polarization reversal in the free ferromagnet{, which results in} a metal-state to insulator-state transition in the device. We also find that, the proposed effect can be realized in devices using other magnetic insulators such as EuO and EuS. Our work should be helpful for developing a realistic switching device that is energy saving and CMOS-technology compatible.
\end{abstract}

%\pacs{ }
\date{\today} %15 February 2017
\maketitle

\section{Introduction}

Modern hard-drive read heads and magnetoresistance random access memories %(MRAM)
are based on a tunneling magnetoresistance effect in a multilayered magnetic tunneling junction structure \cite{wolf2001spintronics,vzutic2004spintronics}.
The structure
comprises one pinned and one free ferromagnets whose relative magnetization
orientations can be switched between parallel and
antiparallel configurations by an external magnetic field,
yielding the desired low- and high-resistance states \cite{chappert2007emergence}.
About two decades ago, a colossal magnetoresistance (CMR) effect
with {a much larger on-off ratio ($\sim 100\times10^3\%$)} \cite{jin1994thousandfold}, %,jin1994colossal}
hence showing the possibility for a ``next generation'' computer hard drive, %\cite{ramirez1997colossal}, %,dagotto2003brief
{was} found in multicomponent manganite perovskites \cite{ramirez1997colossal}. %or manganite perovskites
With decreasing temperature, %proper hole doping
the materials display a transition from a
paramagnetic insulator to a %low-temperature
ferromagnetic half-metal \cite{satpathy1996electronic,park1998direct}. %park1996electronic
Near the transition temperature,
an external magnetic field can drive the insulator
to a quasi-metal \cite{ramirez1997colossal},
yielding %the desired ``giant" or
the ``colossal" on-off ratios.
A strong magnetic field is required for a relatively small low-resistance.

%Very recently, colossal and extremely large magnetoresistance were explored
%in Dirac materials.
%For example,
Several years ago, {a} colossal negative magnetoresistance
%extremely large magnetoresistance
was explored in functionalized graphene %\cite{}
such as
dilute fluorinated graphene \cite{hong2011colossal,Georgakilas2012Functionalization}.
%at extreme low-temperature and high-magnetic field
%The adatom-induced magnetism and/or a metal-insulator transition
%driven by quantum interference may play an important role.
%Hong, X., et al.
%"Colossal negative magnetoresistance in dilute fluorinated graphene."
%Physical Review B 83.8 (2011): 085410.
%Georgakilas, Vasilios, et al.
%"Functionalization of graphene: covalent and non-covalent approaches, derivatives and applications."
%Chem. Rev 112.11 (2012): 6156-6214.
%Recently, giant negative magnetoresistance was recently explored in
%functionalized graphene at extreme low-temperature and high-magnetic field \cite{}.
%two-dimensional materials, such as
Very recently, {an} extremely large magnetoresistance (XMR) was explored in other Dirac materials.
For example, unsaturated XMR up to $0.45\times 10^6\%$ at 4.5 {K} in a magnetic field of 14.7 {T}
and XMR up to $13\times 10^6\%$ at 0.53 {K} in a magnetic field of 60 {T}
{were} observed in WTe$_2$ \cite{ali2014Large},
% 452,700 per cent at 4.5 {K} in a magnetic field of 14.7 teslas, and 13 million per cent at 0.53 {K} in a magnetic field of 60 teslas.
%In contrast with other materials, there is no saturation of the magnetoresistance value even at very high applied fields
XMR of about one million percent at 2K and 9T {was} observed in LaSb \cite{tafti2015resistivity},
XMR of $0.1\times10^6\%$ and $0.73\times10^6\%$ at 2.5 K and 14 T {were} obtained in NbAs$_2$ and TaAs$_2$, {respectively} \cite{wang2016resistivity},
%Here, high quality single crystals of
%NbAs2/TaAs2 with inversion symmetry have been grown, and the resistivity under magnetic field is systematically investigated. Both of them exhibit metallic behavior under zero magnetic field, and a metal-to-insulator transition occurs when a nonzero magnetic field is applied, resulting in XMR (
%1.0��10 5\%  for NbAs2 and 7.3��10 5\% for TaAs2  at 2.5 K and 14 T).
and unsaturated XMR up to $11.2\times 10^6\%$  at 1.8K  in a magnetic field of 33 T {was} obtained in PtBi$_2$ \cite{gao2017extremely}.
% \cite{ali2014Large,wang2016resistivity,tafti2015resistivity,gao2017extremely}
%the physics?
However, to achieve these CMR or XMR's,
extreme high-magnetic {fields} and low-{temperatures} are required.
In these cases, it is not applicable for any realistic devices yet.
%gmr in other ferromagnetic graphene devices?

%Electric-field-induced switching would
%be far more energy saving and compatible with the ubiquitous
%voltage-controlled semiconductor technology \cite{chiba2008magnetization,ohno2010window,matsukura2015control}.
%Successes contain the Datta-Das spin field effect transistor
%in which the spin precession is tuned by a gate voltage
%through a strong spin-orbit coupling \cite{datta1990electronic,nitta1997gate,chuang2015all},
%electric-field-controlled magnetisms in diluted magnetic semiconductors
%in which the ferromagnetism is mediated by carrier \cite{ohno2000electric,chiba2003electrical},
%electric-field-induced magnetization reversals in a multiferroics-based
%heterostructure \cite{garcia2010ferroelectric,heron2011electric},
%and even
%electric-field-induced insulator-metal transitions in magnetoresistive manganites \cite{asamitsu1997current,ponnambalam1999electric}.

\begin{figure}[!t]
  \centering
  \includegraphics[width=0.85\linewidth]{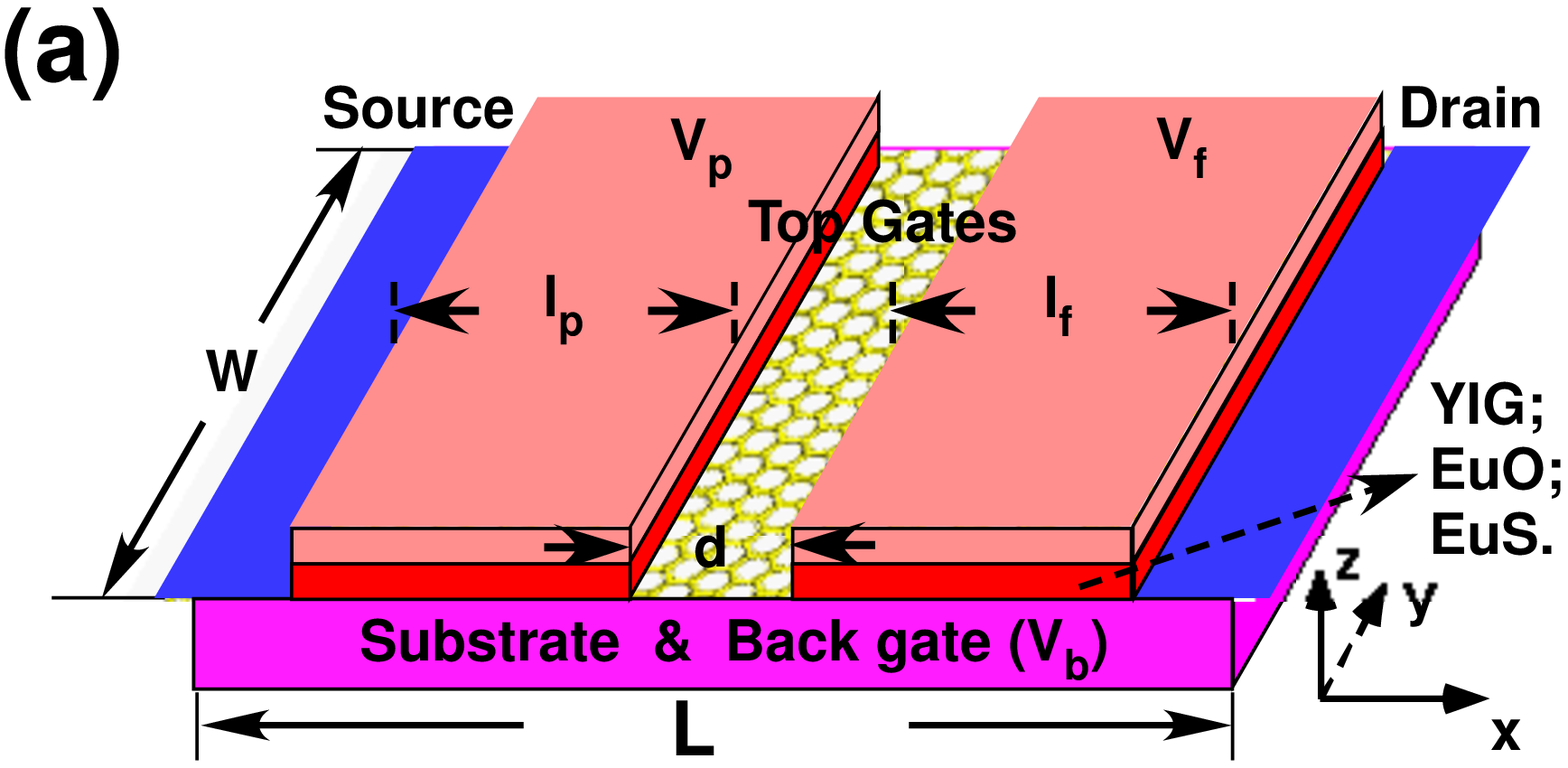}\\
  \includegraphics[width=0.49\linewidth]{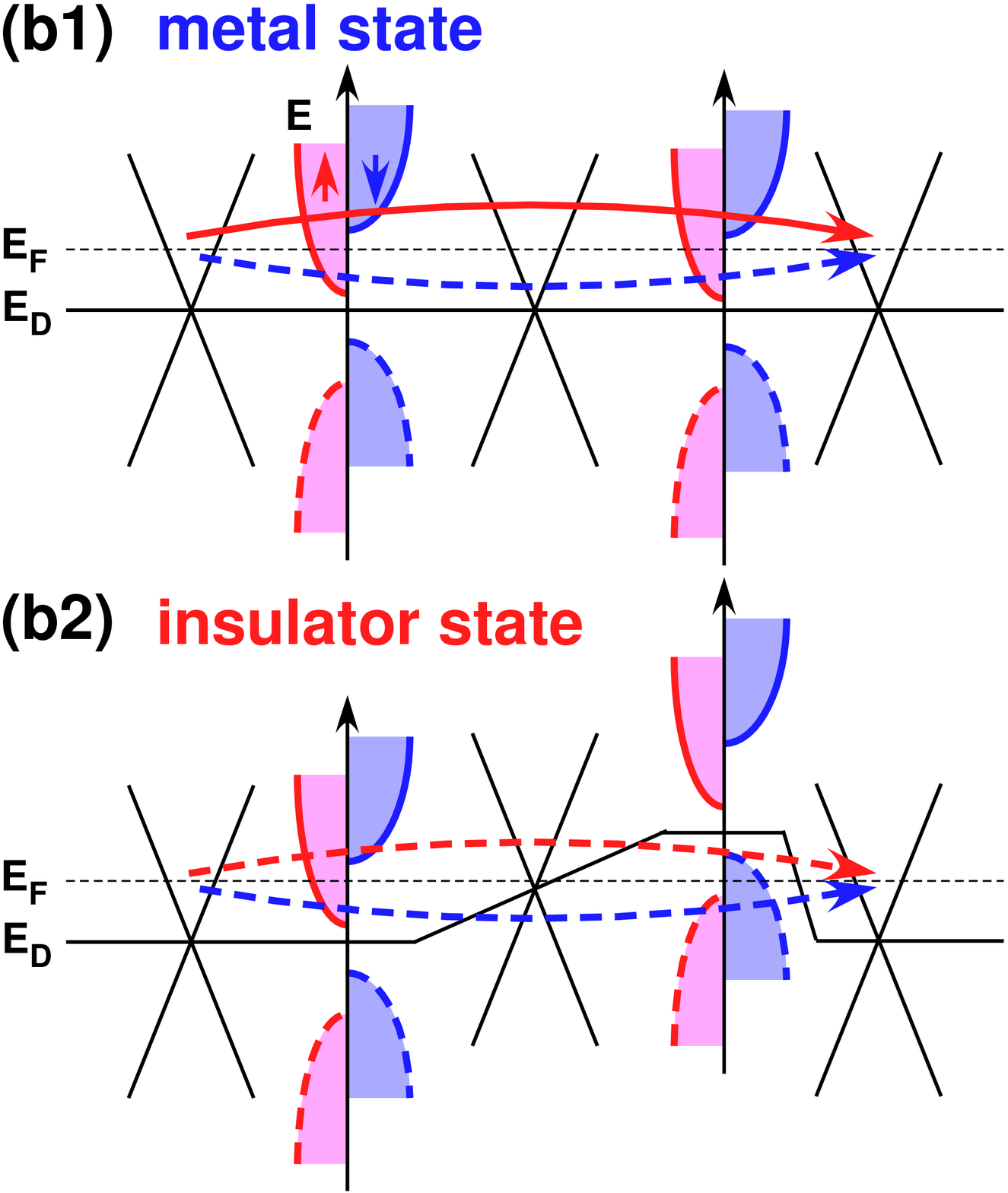}
  \includegraphics[width=0.49\linewidth]{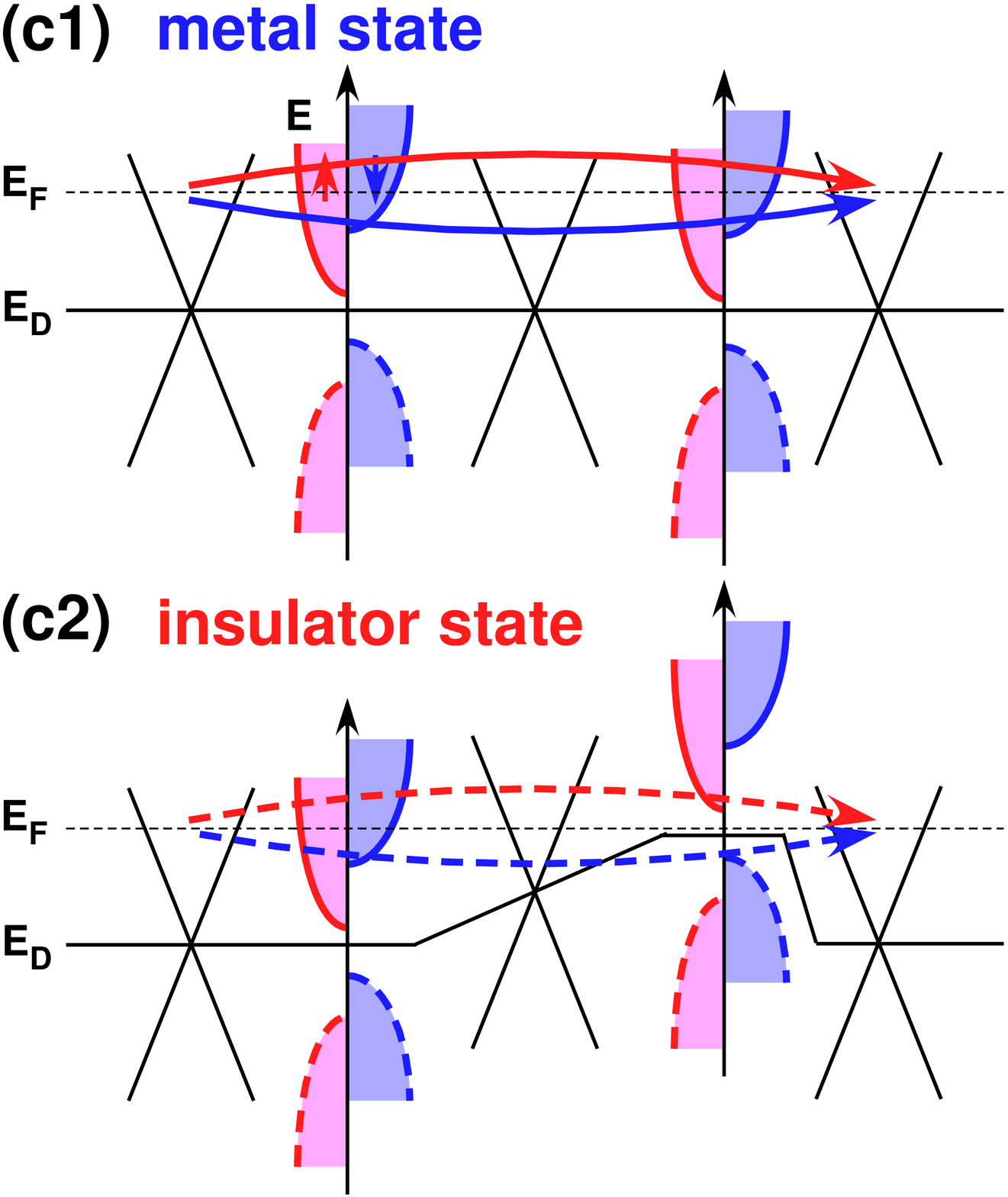}
  \caption{ %the arrows should be from source to drain or opposite!
  Schematic diagrams for (a) the propose device
  and (b,c) its metal and insulator states with half-metallic (b) or normal (c) ferromagnets.
  %(c1) and (c2) schematic diagram for metal state and insulator state of device with normal ferromagnets.
  The ballistic and evanescent transports of each spin are indicated by the solid and dashed lines, respectively.
    }\label{fig:setup-mechanism}
\end{figure}

In this work, we propose a device that can generate %metal-insulator states and
even higher %extremely large
on-off ratios
at {a} high temperature and in {a} zero magnetic field.
%Here we consider an electric-field-induced
%``metal-insulator transition'' in a graphene spin valve, which
The device is composed of {a} graphene under two strips of a magnetic insulator yttrium iron garnet (YIG),
%one of the magnetic insulators,
where {two} gate voltages are applied
to cancel the heavy charge doping's in the YIG-induced half-metallic ferromagnets (see Fig. \ref{fig:setup-mechanism}(a)).
%The Dirac points of the YIG-induced ferromagnets are set to
%align with that of the prinstine graphene and the Fermi energy is tuned into
%%by a back gate,
%the electron exchange splitting window of the ferromagnets?
{Based on calculations using the Landauer-B\"{u}ttiker formalism}, we demonstrate that, by applying a proper voltage on one of the ferromagnets (the free one), %the free electrode,
%the device can change from a metal state %with low resistance (see Fig. \ref{fig:setup-mechanism}(b))
%that decreases with increasing temperature,
%to an insulator state, %with high resistance (see Fig. \ref{fig:setup-mechanism}(c)).
%that decreasese with increasing temperature
%%%%%
%%%%%
%%%%
%Such a change
%resulting in
an extremely large change in resistance
%a extremely large resistance difference,
%which is found high
up to $305\times10^6\%$ %$3.0\times10^8\%$
can be {achieved} at the liquid helium temperature (4.2K) and
in a magnetic filed of 0 T.
The value maintains {as $16\times10^3\%$} at the liquid nitrogen temperature (77K).
%\textcolor[rgb]{1.00,0.00,0.00}{The former (latter) is %comparable with or
%hundreds of times higher than XMR
%previously reported experimental values in any other Dirac materials \cite{ali2014Large,wang2016resistivity,tafti2015resistivity,gao2017extremely}
%(conventional materials
%%under magnetic fields
%\cite{parkin2004giant,yuasa2004giant,chopra2005quantum})
%under magnetic fields of several Teslas.}
%which is hundreds of times higher than
%previously reported experimental values in
%The numerical results
%%show that, %compared with the conventional magnetic-field-induced CMR,
%%the electric-field-induced transition results
%predict extremely large resistance difference up to
%%several $10^6\%$ to
%several $10^8\%$ %$3.0\times10^8\%$, $4.9\times10^6\%$, and $6.5\times10^9\%$
%at the liquid helium temperature,
%%for graphene on YIG, EuO, and EuS, respectively,
%%the latter of
%which is %comparable with or
%hundreds of times higher than
%previously reported experimental values in any other Dirac materials
%under magnetic fields of several Teslas \cite{ali2014Large,wang2016resistivity,tafti2015resistivity,gao2017extremely}.
%For graphene on YIG the numercial results predict value maintains as up to $1.5\times10^4\%$ at the
%liquid nitrogen temperature, which is hundreds of times higher than
%previously reported experimental values in conventional materials
%%under magnetic fields
%\cite{parkin2004giant,yuasa2004giant,chopra2005quantum}.
We indicate that, such {huge on-off ratios}
%the metal-insulator transition
stem from an electric-field-induced reversal of the {full polarization} in the free ferromagnet, %or insulating statte (),
which results in a {transition between a metal and an insulator states}
%ballistic transport of one spin to an evanescent transport of both spins
in the device.
%which ... polarization revision or ferromagnetism-insulation ... .
%which leads to an insulator state in the device
%The insulator state can also be achieved by tuning a normal ferromagnetism to an insulating state in the free electrode.
%The later stems from a full-polarization reversal of half-metallic ferromagnetism
%or ferromagnetism-insulating state .
%the EFE
%half-metallic ferromagnetisms of opposite polarizations (under opposite polarities)
%can be achieved in graphene on some of the magnetic insulators such as YIG and EuO,
%while an insulating state can be achieved in graphene on some others such as EuS and EuO. %\cite{},
%We demonstrate that, %For any high temperature below the Cuire temperature,
%a spin valve with two normal or half-metallic
%ferromagnetisms of a same polarity shows a metal behavior;
%it turns into an insulator behavior
%when the normal ferromagnetism is tuned to an insulating state
%or the polarity hence polarization of the half-metallic ferromagnetism
%is reversed in the free electrode.
%We attribute such a remarkable transition to %an electric-field-induced
%three magnetic insulators which induce different gap and ... are considered.
%all can ... .
%In both cases, a ballistic transport of two or one spins change to an evanescent transport of both spins.
%The proposed mechanism and effect can be explored in other 2D materials,
%provided ferromagnetism (induced or intrinsic) with the desired features
%as well as an EFE exist simultaneously.
Moreover, we find that, the proposed effect {can be} realized
in graphene under two strips of other magnetic insulators, such as
%devices with magnetic insulators of
EuO and EuS.

%\section{Setup and formula}
\section{Device setup}
\begin{figure}[!t]
  \centering
  \includegraphics[width=1.0\linewidth]{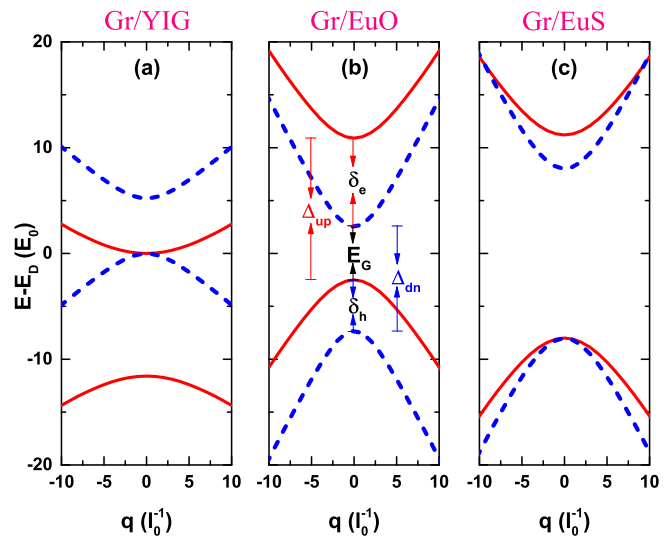}\\
  \caption{Low-energy spin-resolved dispersions %around the Dirac points
  for graphene under (a) YIG, (b) EuO, and (c) EuS;
  solid for spin up and dashed for spin down.
  %The ferromagnetism and insulation windows are labeled in (a) and (b).
  %The Dirac gap ($E_G$), exchange splittings ($\delta_{e,h}$),
  %and spin Dirac gaps ($\Delta_{\uparrow,\downarrow}$) are labeled in (b) for comparison.
 The energy ranges for the hole exchange splitting window, the Dirac gap,
 and the electron exchange splitting window are
  $(-|\delta_h|-E_G/2,-E_G/2)$,    $(-E_G/2,E_G/2)$, and  $(E_G/2,E_G/2+|\delta_e|)$, respectively.
For graphene under YIG, the energy ranges  are $(-11.55E_0,-0.05E_0)$, $(-0.05E_0,0.05E_0)$,
and $(0.05E_0, 5.25E_0)$, respectively.
  For graphene under EuO, they read $(-7.2E_0,-2.5E_0)$, $(-2.5E_0,2.5E_0)$, and $(2.5E_0, 10.9E_0)$, respectively.
 For graphene under EuS, the hole exchange splitting window disappears
 and the energy ranges for the other two windows are $(-8E_0,8E_0)$ and $(8E_0,11.2E_0)${,
 respectively}.
 Here $E_0=10$meV is the energy unit.
  %  this figure and Eqs. 2 and 3 can be used to investigate nanostructure?
}\label{fig:dispersions}
\end{figure}

%Taking these experimental features into account,
%we design a device (why is it a spin valve?) as
The designed device is shown in Fig. \ref{fig:setup-mechanism}(a).
A sufficiently wide (in the {$y$-direction}) and short (in the {$x$-direction})
graphene strip of $L\times W$ is grown on a substrate.
%\textcolor[rgb]{1.00,0.00,0.00}{, which is contacting a back gating ($V_b$)}.
Here, $W$ is several times of $L$ to ensure that the transport is dominated
by bulk states and does not depend on the edge types
\cite{tworzydlo2006sub}.
On top of the graphene, two YIG strips %deposited on GGG
%(or EuO, EuS)
of lengths $l_{p,f}$ and a distance $d$ %, and a width $W$
are transferred \cite{wang2015proximity}. %(or deposited \cite{swartz2012integration,wei2016strong}).
As demonstrated by experiments \cite{wang2015proximity,mendes2015spin,leutenantsmeyer2016proximity,evelt2017chiral}
and {first-principle} calculations \cite{hallal2017tailoring}, the YIG strips induce ferromagnetism in the underlying graphene
through an overlap of the spin-polarized Fe-$3d$ {states} and the C$-p_z$ {states},
which is {called} a magnetic proximity effect.
The {electronic structure of the} graphene ferromagnet can be described by
an exchange splitting for electrons or holes ($\delta_e$ {or} $\delta_h$)
and a band gap opening at the Dirac point ($E_G$),
see Fig. \ref{fig:dispersions}(b).
%The magnetic proximity effect stems from
Unfortunately, %experiments and calculations
%also show that
{a} heavy electron doping ($E_D=-0.78eV$)
%, see Table \ref{parameter})
is also induced in {the graphene ferromagnet} \cite{wang2015proximity,hallal2017tailoring},
which limits its spintronic application by a low polarization.

On the YIG strips two top gates ($V_{p,f}$)
are placed \cite{wang2015proximity}.
%where a fixed and variable voltage is applied.
%to form a pinned and free electrode.
%Here, the magnetic insulators also work as a dielectric \cite{bokdam2013field}.
They are used
to cancel the heavy {doping's} %and tune the dirac point
through a strong {electric-field} effect \cite{wang2015proximity}
similar to that in pristine graphene \cite{novoselov2004electric}.
The substrate under graphene is contacting a back gate ($V_b$), which is used to
tune the Fermi energy ($E_F$) through the whole device.
When the Fermi energy is set as $E_F\in(E_G/2, E_G/2+\delta_e)$ or
$E_F >E_G/2+\delta_e$, {a half-metallic or normal ferromagnet} of electron polarity
{is} made use of.
A small voltage ($\Delta V_f$) is further applied on the free ferromagnet
to tune its Dirac point ($E_{D,f}$). %in the free ferromagnetic barrier.
%therefore, the barrier can be different from the pinned barrier.
As we will see below, %the back gate is used to ... .
the Fermi energy and the Dirac point of the free ferromagnet
%small voltage on the free gate
%difference between the top gates
are important to the states of the device.
%we will see that, by proper Fermi energy and dirac point ...
On the two sides {of the device}, the graphene film is contacted with a source and drain electrode,
which may induce a contact doping ($U$) and a contact resistance \cite{xia2011origins,robinson2011contacting,song2012determination}.
Instead of YIG, EuO and EuS can be deposited on the graphene {film}.
%used as the magnetic insulators;
%the difference from YIG is that,
The ferromagnetism is induced by {an} overlap of {the} Eu$-4f$ states and {the} C$-p_z$ states
and all the parameters ($\delta_{e,h}^{(i)}$, $E_G^{(i)}$, $E_D^{(i)}$) \cite{hallal2017tailoring}
{as well as the Curie temperatures ($T_c^{(i)}$) \cite{wang2015proximity,swartz2012integration,wei2016strong}
become different.
Here $i=1,2,3$ for YIG, EuO, and EuS, respectively.} %, doping, and ...

%The states of the device is controlled by the back gate ($V_b$,
%which determines the Fermi energy) and the top gate
%on the free electrode ($\Delta V_D^f$, which determines the Dirac point).

%In pristine graphene, the Dirac point can be tuned by a gate voltage ($V_g$) following a relation of
%$\Delta E_D \sim \textmd{sign}(V_g)\sqrt{|V_g|}$ \cite{kim2012direct,bokdam2013field}.
%Here the anti-parabolic dependence arises from a linear dispersion of pristine graphene.
%In a ferromagnetic graphene, the dispersions become parabolic (see below); as a result,
%the $E_{D,f}^{(i)}-\Delta V_f$ relation would be changed. %for mp induced ferromagnetic graphene.
%the superscript $i=G,O,S$ stands for YIG, EuO, and EuS, respectively.
%Since the specific relations are still not known, we
%can denote them as $E_{D,f}^{(i)} =\mathcal{F}_\textmd{(i)}(V_f^{(i)})$.
%%where $\textmd{Y,O,S}$ stands for graphene on YIG, EuO, and EuS, respectively.
%To neutralize the heavy electron doping induced by the magnetic insulators, % \sim 1eV
%see Fig. \ref{fig:setup-mechanism} (b1),
%the top gates are set as $V_{p,f}=\mathcal{F}^{-1}(-E_D)$.

\begin{table}[h]
\small
  \caption{The Curie temperatures, Dirac cone doping's (in unit of eV),
  spin Dirac doping's (in unit of $E_0$), spin Dirac gaps (in unit of $E_0$),
and spin Fermi velocities (in unit of $v_F$) for graphene under YIG, EuO, and EuS. }
  \label{parameter}
  \begin{tabular*}{0.48\textwidth}{@{\extracolsep{\fill}}lllcccccccc}
    \hline
   & $T_c$ & $-E_D$ & $D_\uparrow$ & $\Delta_\uparrow$ & $v_\uparrow$ & $D_\downarrow$ & $\Delta_\downarrow$ & $v_\downarrow$  \\
    \hline
    YIG/Gr & 550K %\cite{wang2015proximity}
& 0.78 & -5.8 & 11.6 & 0.63
& 2.6 & 5.3 & 0.70 \\ %-780-58 26
EuO/Gr & 69K %\cite{swartz2012integration}
& 1.37 & 4.2 & 13.4 & 1.337
& -2.4 & 9.8 & 1.628 \\ %-1370+84/2 -48/2
EuS/Gr & 16.5K %\cite{wei2016strong}
& 1.3 & 1.6 & 18.3 & 1.40
& 0 & 15 & 1.70 \\ %-1300+32/2 0/2
    \hline
  \end{tabular*}
\end{table}

\section{calculation formula}
Below we will derive the formula to calculate the device conductance
at different $E_F$ and $E_{D,f}$.
%and spin-resolved.
Devices using YIG, EuO, and EuS will be handled uniformly.

The low-energy dispersions around the Dirac points, which are
predicted by first-principle calculations \cite{hallal2017tailoring,yang2013proximity},
are re-calculated and plotted in Fig. \ref{fig:dispersions}.
The dispersions {read} %and parameters.
\begin{equation}\label{Eq:dispersion}
  E_s^{(i)}(q)-E_D^{(i)} = D_s^{(i)}\pm\sqrt{(\hbar v_s^{(i)} q)^2+(\Delta_s^{(i)}/2)^2}.
\end{equation}
{Here} $q$ is the momentum, $D_s^{(i)}$, $v_s^{(i)}$, and $\Delta_s^{(i)}$
($s=\pm1$ for spin up and down) are the material- and spin-dependent
Dirac cone dopings, Dirac gaps, and Fermi velocities, respectively.
From {Fig. \ref{fig:dispersions}}, one can see clear spin-resolved
parabolic dispersions, {which} accompany with a band gap opening at the Dirac point %($E_G^{(i)}$)
and {exchange splitting's for electrons and holes}. %($\delta_e^{(i)}$ and $\delta_h^{(i)}$).
The {parameters} for graphene on six trilayer YIG (six bilayer EuO and EuS) are given in {Ref.} \cite{hallal2017tailoring}.
%Here are considered.
The parameters in Eq. (\ref{Eq:dispersion})
relate with them by
%$E_G^{(i)}$ and $\delta_{e,h}^{(i)}$ \cite{hallal2017tailoring} by
$D_{\uparrow,\downarrow}^{(i)}=\pm\delta_{e,h}^{(i)}/2$ and %$E_{D\downarrow}=E_D-\delta_h/2$,
$\Delta_{\uparrow,\downarrow}^{(i)}=|\delta_{e,h}^{(i)}|+E_G^{(i)}$,
see Fig. \ref{fig:dispersions}(b);
$v_{\uparrow\downarrow}$ are fitted from the original dispersions \cite{hallal2017tailoring}.
{The values} are listed in table \ref{parameter}.

Regarding Eq. (\ref{Eq:dispersion}) as a combination of dispersions of two gapped graphene,
the dispersions of the graphene ferromagnets can be described by an uniform effective Hamiltonian
in a sublattice space \cite{beenakker2008colloquium}
\begin{equation}\label{Hamiltonian}
\mathcal{H}_{\bm{k},s,\xi}^{(i)}=\mathcal{I} (D_s^{(i)} + E_{D,f}^{(i)}) + \xi \sigma_z \Delta_s^{(i)} + \bm\sigma \cdot \hbar v_s^{(i)} \bm{k},
\end{equation}
where
$\mathcal{I}$ is the identity matrix,
$\xi =\pm 1$ is the index for valley $K$ and $K^\prime$,
$\bm{\sigma}=(\sigma_{x},\sigma _{y})$ is the pseudospin Pauli matrices,
and $\bm{k}=(k, q)$ is the momentum operator.
{Such an} effective Hamiltonian is different from those described
in a sublattice-spin direct produce space \cite{zollner2016theory,su2017effect,hallal2017tailoring},
{where} the ferromagnets are viewed as
a Dirac gap and two exchange splittings.
Note, in this space a four-component wave function should be solved \cite{song2013ballistic}.
Different from them,
in Eq. (\ref{Hamiltonian}) we also consider a spin-resolved Fermi velocity $v_s$.
As we will show below, this parameter would determine an effective spin Dirac gap ($\Delta_s/v_s$) hence
{play} an important role in the insulator state. %and change in resistance. %and CMR.
%%%
For the graphene under the electrode metals,
$\mathcal{H}_{\bm{k},\xi}=\mathcal{I} U + \bm\sigma \cdot \hbar v_F \bm{k}$,
and for the pristine graphene between the pinned and free ferromagnets,
$\mathcal{H}_{\bm{k},\xi}=\mathcal{I} \phi(x) + \bm\sigma \cdot \hbar v_F \bm{k}$,
where $\phi(x)=(x/d) E_{D,f}$ is the potential shift.

%%%%
When $E_{D,f}^{(i)}=0$ (the cases in Fig. \ref{fig:setup-mechanism} (b1) and (c1)),
the right- and left-going envelope functions in
the contacted, ferromagnetic, and pristine
graphene ($j=c, m, p$) %song2015spin,
can be obtained by exactly resolved
$\mathcal{H}_j\Phi_j^\pm=E_j\Phi_j^\pm$. %\cite{song2012giant,song2013generation,song2015spin,song2016spin}.
Using a characteristic energy (length) $E_0=10$ meV ($l_0=\hbar v_F/E_0=56.55$ nm),
the dimensionless result reads
%\begin{equation}
%\Phi_j^\pm=[1,(\pm k_j+iq_j)/E_j]^T e^{\pm ik_jx+iq_jy}/\sqrt{2},
%\end{equation}
\begin{equation}\label{wave}
%\boldsymbol{\Psi}
\boldsymbol\Phi_j^\pm=\frac{1}{\sqrt{2}E_j}
\left(\begin{array}{c}
E_j\\
\pm k_j+iq_j\end{array}\right)e^{\pm ik_jx+iq_jy},
%+q\left(\begin{array}{c}
%1\\
%(- k_j+iq_j)/E_j\end{array}\right)\frac{e^{- ik_jx+iq_jy}}{\sqrt{2}},
\end{equation}
where $E_{p(c)}=E(-U)$,
$E_m=(E_s+\Delta_s)/v_s$ with $E_s=E-D_s(-E_{D,f})$ for the pinned (free) ferromagnet,
$q_{p,c,m}=E_{c}\sin\alpha$ is the conserved transverse wave vector,
$k_{p,c}=\textmd{sign}(E_{p,c})(E_{p,c}^{2}-q_{p,c}^{2})^{1/2}$, and
$k_m=\textmd{sign}(E_m)(E_m E_m^\prime-q_m^{2})^{1/2}$ with $E_m^\prime=(E_s-\Delta_s)/v_s$.
It is seen that,
%At the spin Dirac point ($E_s=0$),
$k_m^2+q_m^2=(E_s^2-\Delta_s^2)/v_s^2$.
{This means that,} an effective spin Dirac gap, $\Delta_s/v_s$, is determined by
not only the spin-dependent Dirac gap but also the spin-dependent Fermi velocity.

%%%
When $E_{D,f}^{(i)}>0$ (the cases in Fig. \ref{fig:setup-mechanism} (b2) and (c2)),
%The difference in Dirac points in the two electrodes also induces %an electric field hence
%a voltage shift hence an n-p (p-n) junction between the free and pinned (drain) electrodes,
%see Fig. \ref{device} (b2).
the graphene between the ferromagnets %free ferromagnetic electrode and pinned electrode (source electrode)
becomes an n-p junction and the envelope function cannot
be straightforwardly solved \cite{sonin2009effect,song2013negative}.
Instead, the function can be solved in a pseudospin space which is rotated
by $\pi/2$ around the $y$-axis (see Fig. \ref{fig:setup-mechanism}(a)) \cite{sonin2009effect,song2013negative}.
The result reads
\begin{equation}
\boldsymbol\Phi_p=c^+\left(\begin{array}{c}
F(x)\\
G^{*}(x)\end{array}\right)e^{i q_p y}
+c^-\left(\begin{array}{c}
G(x)\\
F^{*}(x)\end{array}\right)e^{i q_p y},
\end{equation}
where $F(x)=D[-1+iq^2/2f,(1+i)(E+fx)/\sqrt{f}]$ and
$G(x)=(1+i)\sqrt{f}q^{-1}D[iq^2/2f,(1+i)(E+fx)/\sqrt{f}]$
with $D[,]$ being the Weber parabolic cylinder function %$q^2=k_y^2+\delta^2$
and $f=E_{D,f}/d$. %($L=2l_b+l_1+d+l_2<L_t$ is the total length between the source and drain).
Note, $F(x)$ and $G(x)$ [$F^{*}(x)$ and $G^{*}(x)$] have {the} properties of a right
(left)-going evanescent wave function \cite{sonin2009effect}. %\cite{rotation}.
In the rotated pseudospin space, the envelope functions for the contacted and ferromagnetic graphene
become $\boldsymbol\Phi_j^\pm=[E_j\pm k_j+iq_j,-E_j\pm k_j+iq_j]^T(2E_j)^{-1} e^{\pm ik_jx+iq_jy}$.

For $T<\min(100K, T_c^{(i)})$, the ferromagnetisms hold and {the}
inelastic scatterings ($\textmd{e}$-$\textmd{e}$ and $\textmd{e}$-$\textmd{ph}$)
can be ignored \cite{morozov2008giant,chen2008intrinsic}.
The spin-resolved conductance through the device can be given
by the Landauer-B\"{u}ttiker formula \cite{buttiker1985generalized}
\begin{equation}\label{T-current}
\begin{aligned}
G_s^{(i)}(E& _{D,f},E_F,T)=\\
G_{0}&\int dE\frac{-df(E,T)}{dE}\int_{-\pi /2}^{\pi /2}|t_s^{(i)}(E_{D,f},E,\alpha )|^2\cos \alpha d\alpha,
\end{aligned}
\end{equation}%
where
$G_{0}=2 e^{2}/h\times(|E_{F}|W/h v_F)$ is a unit conductance,
$f(E,T)=[1+e^{(E-E_{F})/k_BT}]^{-1}$ is a Fermi-Dirac distribution function,
and $t_s(E_{D,f},E,\alpha)$ is a spin-resolved transmission coefficient
at a Dirac point of $E_{D,f}$, an energy of $E$, and an incident angle of $\alpha$.
$t_s^{(i)}(E_{D,f},E,\alpha)$ can be solved by $\boldsymbol\Phi_j$ with a standard transfer matrix method \cite{born1980principles}.
The $E_{D,f}$-dependent resistance of the device can be given by
$R(E_{D,f},E_F,T)=[G_\uparrow(E_{D,f},E_F,T)+G_\downarrow(E_{D,f},E_F,T)]^{-1}$ in unit of $R_0=G_0^{-1}$.
%As we will see below, the resistance of the device will change dramatically
%with $\Delta E_D$,
The {changes} in resistance (RC) at different $E_{D,f}$ can be defined as
%The CMR is defined as
\begin{equation}
\begin{aligned}
\textmd{RC}_{(i)}& (E_{D,f}^{(i)},E_F,T)=\\
& \frac{R_I(E_{D,f}^{(i)},E_F,T)-R_M(0,E_F,T)}{R_M(0,E_F,T)}\times 100\%{.}
\end{aligned}
\end{equation}
%where $R_M$ and $R_I$ are a resistance for a metal or insulator state
%at zero or a certain $\Delta E_D$.
To calculate a change in resistance, four conductance, {
$G_{\uparrow,\downarrow}(0,E_F,T)$ and %, $G_\downarrow(0,E_F,T)$
$G_{\uparrow,\downarrow}(E_{D,f},E_F,T)$, %, and $G_\downarrow(\Delta E_D,E_F,T)$
should be calculated.}
Each of them depends strongly on the magnetic insulator ($i$),
ferromagnet length ($l_{p,f}$), electrode contact ($U$), Fermi energy ($E_F$), and temperature ($T$).

\begin{figure*}[!t]
  %\flushleft
  \centering
  \includegraphics[width=\linewidth]{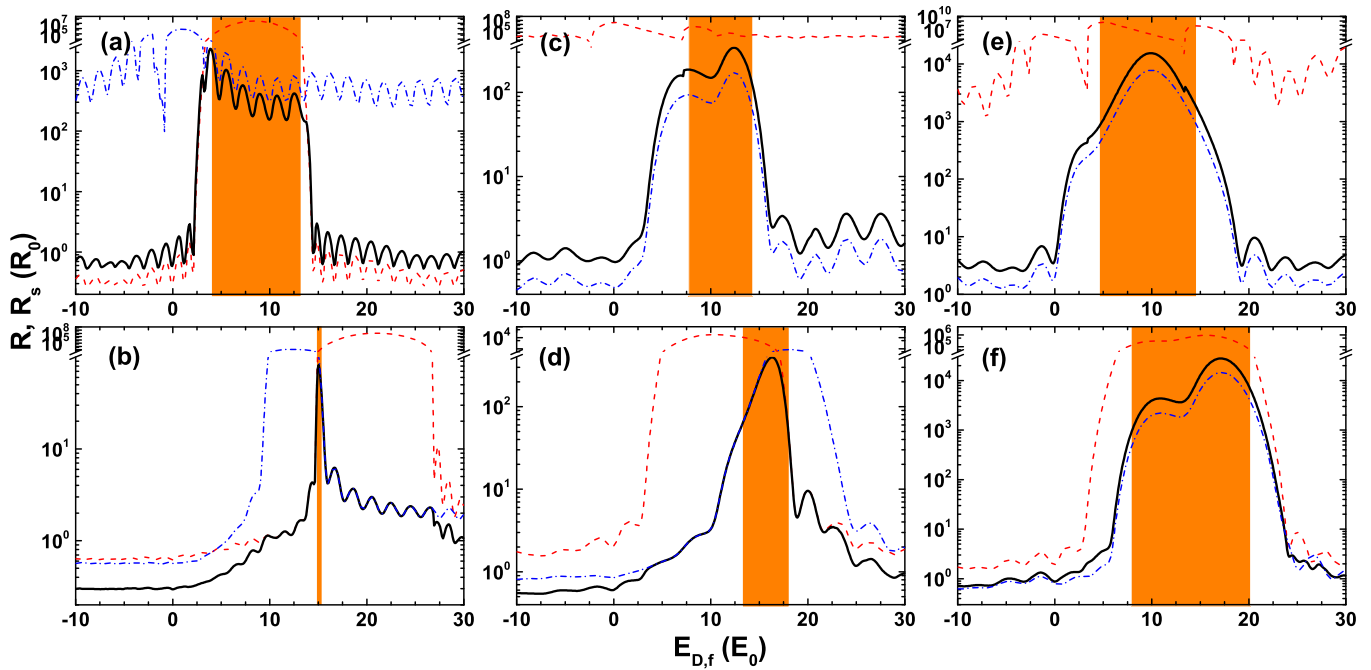}\\
  \caption{
  The device resistance and its spin components (red dashed for spin up and blue dashed for spin down)  as a function of the Dirac point shift in the free ferromagnet. %$V_f-V_p$
  The parameters are $l_p$=$l_f$=1.0, $d$=0.5, $U=0$, and $E_{D,p}=0$.
  (a, c, e) The cases for devices with half-metallic ferromagnets from YIG, EuO, and EuS ($E_F=2.5, 7, 9.5$), respectively.
  The {small (large)} spin component is reduced (enlarged) by 2 times for clearness.
  (b, d, f) The cases for devices with normal ferromagnets from  YIG, EuO, and EuS ($E_F=15$), respectively.
   }\label{fig:gate-voltage}
\end{figure*}

\section{Results and discussion}

%since the ferromagnetism of the barriers alter the spins differently,
%we will consider the half-metallic and normal ferromagnetism seperately.

\subsection{Gate-induced huge changes in resistance based on metal-insulator states}

\begin{figure}[!t]
  \centering
  \includegraphics[width=\linewidth]{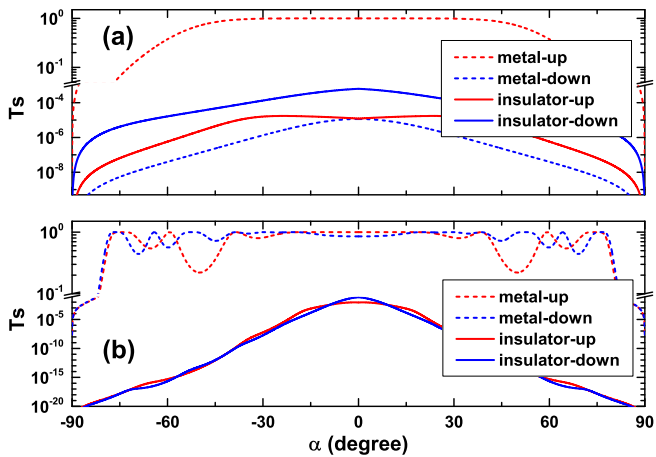}
  \caption{
 % (a) The graphene spin valve based on a combination of
%  a magnetic proximity effect and an electric field effect.
  (a) The spin{-dependent} transmissions for the metal (dashed)
  and insulator (solid) states based on half-metallic ferromagnet{s}.
Red for spin up and blue for spin down.
  (b) The  spin{-dependent} transmissions for the metal (dashed)
  and insulator (solid) states based on normal ferromagnets.
      }\label{fig:transmission}
\end{figure}

\begin{figure}[!t]
  %\flushleft
  \centering
  \includegraphics[width=\linewidth]{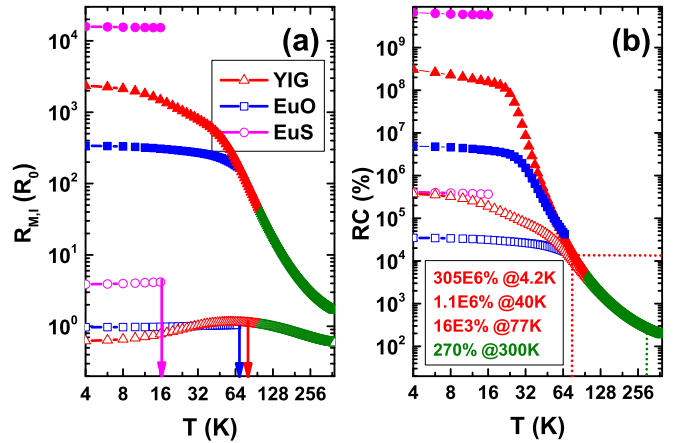}
    \caption{
    (a) The {resistances of} metal (solid) and insulator (hollow) states for devices of $l=1$ based on YIG (red), EuO (blue), and EuS (magenta)  as a function of temperature.
(b) The change in resistance for devices of $l=2$ (solid) and $l=1$ (dashed) based on YIG (red), EuO (blue), and EuS (magenta)  as a function of temperature.
    %in b, add the case for l=2.
%    (b) The corresponding
    %, the cases in Fig. \ref{gates} (a-c).
    %(b) Zero temperature $G_M$ (dashed) and $G_I$ (solid) as a function of energy
    %for spin valves based on YIG (b1), EuO (b2), and EuS (b3). %, the cases in Fig. \ref{gates} (a-c).
    }\label{Fig:temperature}
\end{figure}

We first consider a YIG-based device with two half-metallic ferromagnets ($E_F=2.5$) of $l=1$.
Fig. \ref{fig:gate-voltage}(a) shows the {zero-temperature device} resistance %and its spin components ($R_{\uparrow,\downarrow}=G_{\uparrow,\downarrow}^{-1}$)
as a function of $E_{D,f}$.
%Here,, $U=0$, and $T=0$ is considered.
%We first consider device with half-metallic barriers.
%which reads ..., respectively.
% perhaps the reviewer dose not know how the resistance is exactly calcualted.
%for yig, euo, and eus, the fermi energy is choose as 2.5, 7, and 9, respectively.
% with half-metallic electrodes
% YIG (),%Here half-metallic ferromagnetism of graphene under YIG, EuO
%($E_F=2.5$ and 7) %at is considered.
%are considered.
%We first consider
%or $V_b\in(\mathcal{F}^{-1}(E_G/2), \mathcal{F}^{-1}(E_G/2+\delta_e))$,
%where a normal or half-metallic ferromagnetism (of electron polarity) can be made use of,
%see mode I and mode II in Fig. \ref{device} (b1), respectively.
%The resistance %and its spin components ($R_{\uparrow,\downarrow}=G_{\uparrow,\downarrow}^{-1}$)
%as a function of $E_D^f$
%are plotted in Fig. \ref{fig:gate-voltage} (a-c) for YIG, EuO, and eus, respectively,
%which are exactly calculated from Eq. (\ref{T-current}).
%Let us first consider the Gr/YIG case Fig. \ref{fig:gate-voltage} (a).
As can be seen, the device resistance is rather low ($\sim 10^0R_0$) at $E_{D,f}=0$;
it becomes rather high ($\sim 10^3R_0$) as $E_{D,f}$ enters into the orange window.
This leads to a huge
%The metal-insulator states result in an
on-off ratio of $3.92\times10^5\%$ at $E_{D,f}=3.85$.
%it is surprising
At {$E_{D,f}=0$}, both ferromagnets are half-metallic ({full polarizations of spin-up}). %electron polarity.
%spin what.
Accordingly, electrons of spin up from the device source {are} transparent,
while electrons of spin down {are} blocked by two potential barriers
(see Fig. \ref{fig:setup-mechanism}(b1)).
%They work as barriers for spin down but ... for spin up.
%the fermi energy is higher than the barrier in both electrodes.
{These behaviors result} in a near-unit transmission for spin up and
a rather small transmission ($\sim 10^{-6}$) for spin down, respectively,
see the dashed lines in Fig. \ref{fig:transmission} (a).
%As a result, spin up transports ballistically in the device,
%reflected by the ... transmission plotted in Fig. \ref{fig:transmission}.
The transmissions {respectively} contribute a rather small ($0.62 R_0$) and large ($1.5\times10^5 R_0$) resistance,
see dashed lines in Fig. \ref{fig:gate-voltage}(a).
%while spin down are blocked (transports evanescently through) two barriers,
% imagnary in both ... . (see Mode I in Fig. \ref{mechanism}(a))
%resulting .... in Fig. \ref{fig:gate-voltage}
%see the ... and rather small transmission plotted in
%(The huge difference between the spin resistances also reflects the half-metallic ferromagnetsim.)
%These are clearly seen by
%and the spin compents shown in Fig. \ref{fig:gate-voltage}.
%they are reflected by, respectively.
The total resistance, as a parallel connection of them, %a small and a large one,
is rather small ($0.62 R_0$, almost equals to the small one).
On the other hand, as can be seen in Fig. \ref{Fig:temperature} (a),
the low resistance increases with an increasing temperature, %(see Fig. \ref{temperature} (a)) .
%while the latter decreases with an increasing temperature.
%These behaviors
{which implies} that the device supports a metal state at $E_{D,f}=0$.
%and insulator state
% is formed in the device
%when %behaves as a metal and an insulator at
%zero and finite gate voltage is applied, respectively.

In the orange window, $E_{D,f}\in(3.0,13.8)$ and {
$E_F-E_{D,f}\in(-11.3,-0.5)\in(-|\delta_h|-E_G/2,-E_G/2)$, see Fig. \ref{fig:dispersions}.
The latter %is included by the hole %and electron
%exchange splitting window.
%$=(-11.55,-0.05)$. %and $(E_G/2,E_G/2+|\delta_e|)=(0.05,5.25)$,
%We find that
%which
means that,} %This corresponds to the case that
the Fermi energy lies in the hole exchange splitting window of the free ferromagnet,
see Fig. \ref{fig:setup-mechanism}(b2).
%dominated by the smaller one, %and is smaller than the s.
%%%
%Further more, an insulating state, half-metallic ferromagnetism of hole polarity,
%and normal ferromagnetism of hole polarity
%can be achieved in the free electrode %can be turned to
%by shifting the Dirac point %in the free electrode,
%$E_D^f\in(E_F-E_G/2,E_F+E_G/2)$, $E_D^f\in(E_F+E_G/2,E_F+E_G/2+\delta_h)$,
%and $E_D^f>E_F+E_G/2+\delta_h$, respectively.
%This can be ... through a negative gate on the free electrode ($\Delta V_f=\mathcal{F}^{-1}(-E_D^f)$)
%which induces a light hole doping.
%Considering the energy range (the question ? by reviewer 1),
Importantly, the reversal of the charge polarity leads to an reversal of the spin polarization.
In other words, by a proper $E_{D,f}$ the spin-up full polarization in the free ferromagnet
becomes a spin-down full polarization.
As a result, electrons of spin up from the device source, which are transparent before, now {encounter} a barrier in the free ferromagnet,
%... the transmission is selective.
{On the other hand,} electrons of spin down from the device source, which are blocked by two barriers before, now {are} still blocked by a barrier
in the pinned ferromagnet (see Fig. \ref{fig:setup-mechanism}(b2)).
%and a p-n junction. %and transports evanescently,
%because the wavevector km becomes imaginary,
%For spin down, although the barrier in the free electrode disappears
%the one still exists in the pinned electrode, which leads to ...
%while spin down keep evanesently, as still one of is imaginary.
{These behaviors are} clearly reflected by the extremely suppressed transmissions ($\sim 10^{-5}$)
and rather high resistances ($4.8\times10^4R_0$ and $2.4\times10^3R_0$) for both spins, as shown
%() of both spins
in Fig. \ref{fig:transmission} (a) and Fig. \ref{fig:gate-voltage}(a).
The total resistance, as a parallel connection of two huge spin resistances{, becomes} rather high ($2.3\times10^3R_0$).
Besides the electric-field induced full-polarization reversal,
the electric-field induced n-p junction also contributes to the high resistance state.
{This is because the transmissions become selective} \cite{cheianov2006selective,cheianov2007focusing}.
In Fig. \ref{Fig:temperature} (a), we find that the high resistance decreases
with an increasing temperature, which implies that the device {now supports}
an insulator state.
%(the transmission should be plotted, which is not hard, for all cases
%disscussed above; the selective transmission reflected by compare with
%a step-like voltage?)
%(The selective transmission through a p-n junction \cite{cheianov2006selective,cheianov2007focusing}
%is also helpful for the insulator state.)
%We find that, %for the latter case
%within the window the half-metallic ferromagnetism
%of electron polarity and spin up full polarization in the free electrode becomes
%that of hole polarity and spin down full polarization (see Fig. \ref{Hamiltonian} (a)).
%As a result, spin up transports evanescently in the free electrode
%while spin down transports evanescently in the pinned electrode,
%see ``Mode I'' in Fig. \ref{device} (b2).
%Note, spin up transports ballistically for the metal case, see ``Mode I'' in Fig. \ref{device} (b1).
%besides the change of the free electrode,
%the n-p junction region induced by the ...,
%also contributes to the ... transport, %see Fig. \ref{fig:transmission}.
%since the transmission is selective \cite{cheianov2006selective,cheianov2007focusing}
%in a pure n-p junction structure.

The {resistances} %and its spin components %($R_{\uparrow,\downarrow}=G_{\uparrow,\downarrow}^{-1}$)
of the same device %but with normal ferrromagnetic electrodes %are considered.
%The same device
%We now consider device with
but with normal ferromagnets ($E_F=15$)
%it can be realized by setting the back gate as $V_b >\mathcal{F}^{-1}(E_G/2+\delta_e$).
are plotted in Fig. \ref{fig:gate-voltage} (b) as a function of $E_{D,f}$.
%the cases for YIG, EuO, and EuS are respectively shown in (d-f).
%%for normal ferromagnetism.
%which are exactly calculated from Eq. (\ref{T-current})
%We now consider CMR with a gate-induced insulating state.
%Fig. \ref{gates} (d) shows a case of normal ferromagnetism for graphene under YIG ($E_F=15$), which
%we first consider the Gr/YIG case.
Similar metal state at zero voltage and insulator state at a proper voltage are found.
The resulting on-off ratio reads $1.7\times 10^4\%$.
However, it is noticed that, the $E_{D,f}$ range for the insulator state
is different from {that} in Fig. \ref{fig:gate-voltage}(a)
and it becomes
%happens at a different voltage
%and the voltage range becomes
much narrower. %than the previous one.
In the {orange} window, it is found that $E_F-E_{D,f}\in(-E_G/2,+E_G/2)$
($E_{D,f}\approx15.05$), which means that, the Fermi energy {enters} into the Dirac gap
of the free ferromagnet.
As a result, electrons of both spins from the device source {encounter potential} barriers in the free ferromagnet,
see Fig. \ref{fig:setup-mechanism}(c2).
This is clearly reflected by the rather small {transmissions ($<10^{-4}$)
and the rather high resistances ($381R_0$ and $105R_0$) for both spins,} see Fig. \ref{fig:transmission}(b)
and Fig. \ref{fig:gate-voltage}(b), respectively.
%they transport evanscently and high resistance in.
{The case} is rather different from the zero voltage case, for which
both spins transport over barriers (see Fig. \ref{fig:setup-mechanism}(c1)) and
{result} in near-unit transmissions in Fig. \ref{fig:transmission}(b)
and small resistances ($0.62R_0$ and $0.57R_0$) in Fig. \ref{fig:gate-voltage}(b).
%these ... are ... by the spin components of resistance,
%which are also plotted in Fig. \ref{fig:gate-voltage}
%The ferromagnetism is clearly reflected by the
%low resistances for both spins at zero $E_D^f$.
%One can see that, the spin valve behaves as a metal for $E_D^f=0$ and
%as an insulator for $E_D^f$ entering into the rather narrow orange window.
%This is because of not only the rather low or high resistance but also the
%increasing or decreasing temperature behaviors (not shown).
%Actually, for the metal state at $E_D^f=0$ both spins transport
%ballistically (see ``Mode II'' in Fig. \ref{device} (b1)).
%in the window, the Fermi energy enters into the Dirac gap (see Fig. \ref{Hamiltonian} (a)) , which means that the ferromagnetic graphene in the free electrode becomes an insulating state.
%In this case, longitude wavevectors for both spins become imaginary (see ``Mode II'' in Fig. \ref{mechanism} (b))

From the above results and {discussions} we can see that, colossal changes in resistance,
which require a strong magnetic field in manganite perovskites, can be realized
by a small gate voltage in the proposed device.
%In previous systems,.
The underlying mechanism is that, the gate induces %shift of dirac point,
%comparing the metal and insulator states at different gates,
%it is seen that, the state transition and resistance difference stems from
%Hence, we can conclude that it is
either a full-polarization reversal in a half-metallic ferromagnet
or an energy gap in a normal ferromagnet,
which results in a metal-insulator state of the device.
%change from a ballistic transport of one (two) spin to an evanescent transport of both spins.
%and results in the metal-insulator transition.
%the above mechanism is rather similar to the two currents model.
For the former case, the insulator state arises because
different spins are blocked in different ferromagnets,
%the evanescent transports for different spin happens in the fixed and free electrode, respectively,
while for the latter case, both spins are blocked
%The difference is that the evanescent transport of both spins here
in the free ferromagnet.
%in this view, the normal-insulating state configuration can also be regarded as a spin valve.
%the mechanism is rather similar to that based on half-metallic ferromagnetisms, ``Mode I'' .
%the curie temperature for yig is the highest. however,
The energy gap (1meV) %in graphene under YIG with the highest $T_c$
is rather narrow and will be measured out at high temperature.
In following we will focus on metal-insulator states
stemming from %(finite-temperature) CMR resulting from
the full-polarization reversal. %(except the EuS case).
%The metal and insulator states will be emphasized.

Can the remarkable effect be realized in other systems?
In Fig. \ref{fig:gate-voltage}(c-f) we plot the calculation results for devices
with ferromagnets induced by EuO and EuS.
The half-metallic ferromagnet cases {based on} EuO ($E_F=7$) and EuS ($E_F=9.5$) are shown in Fig. \ref{fig:gate-voltage}(c) and (e),
and the normal ferromagnet cases {based on} them ($E_F=15$) are shown in Fig. \ref{fig:gate-voltage}(d) and (f), respectively.
It is seen that, the device resistance shows similar {dependence on
the Dirac point of the free ferromagnet}
as the YIG cases.
The resulting CMR are $3.5\times10^4\%$, $6.53\times 10^4\%$,
$4.1\times10^5\%$,
%becomes even larger, which reads %$1.7\times 10^4\%$,
and $3.18\times 10^6\%$ for Figs. \ref{fig:gate-voltage}(c-f), respectively.
%394.266 at 16.35 vs. 0.603 at zero
%27869 at 17.15 vs. 0.876 at zero
Meanwhile, the low and high resistance at zero and proper $E_{D,f}$
show similar temperature dependence as the YIG cases (see Fig. \ref{Fig:temperature}(a)).
The above behaviors mean that the voltage-induced metal-insulator states and huge changes in resistance
can also be realized in devices based on other magnetic insulators.

\subsection{Ferromagnet-length dependence of the on-off ratios}

\begin{figure}[!t]
  %\flushleft
  \centering
  \includegraphics[width=\linewidth]{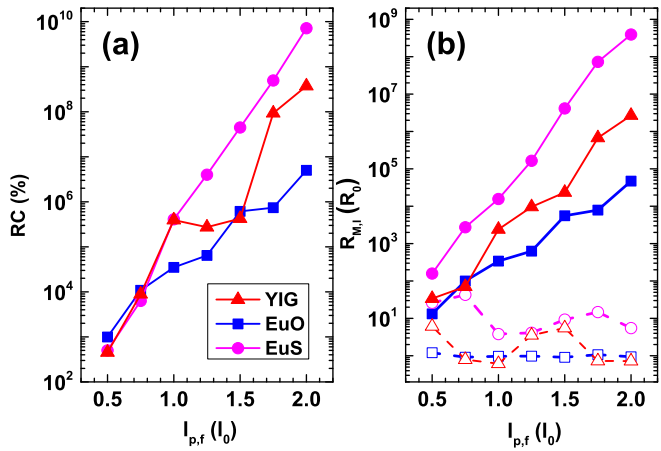}
\caption{(a) The change in resistance as a function of the ferromagnet lengths for
devices based on
%   Metal-like low resistance (hollow), insulator-like high resistance (solid), and CMR
%    as a function of (a) temperature, (c) electrode length, and (d) contact doping
     YIG (red), EuO (blue), and EuS (magenta). %, the cases in Fig. \ref{gates} (a-c).
    %(b) Zero temperature $G_M$ (dashed) and $G_I$ (solid) as a function of energy
    %for spin valves based on YIG (b1), EuO (b2), and EuS (b3). %, the cases in Fig. \ref{gates} (a-c).
(b) The corresponding on-state (dashed) and off-state (solid) resistances.
    }\label{fig:length}
\end{figure}

In all the above calculations, the ferromagnet length is fixed as
%All the above ferromagnets length are
$l_0$; what will happen when it is changed?
In Fig. \ref{fig:length}(a) we plot the on-off ratio %and corresponding metal-insulator states
of {the} YIG-based device as a function of the lengths of the pinned and free ferromagnets.
%The results are calculated by changing $l$ in Eq. \ref{} and fixing others.
%... and plotted in ... .
Surprisingly, it is found that, the on-off ratio is rather sensitive to the {lengths};
%As can be seen,
%Accordingly,
it shows a near-exponential dependence on the ferromagnet lengths
excepting %the cases for graphene under YIG with
1$<l_{p,f}<$1.5. %see Fig. \ref{fig:length} (c2),
When the ferromagnet {lengths increase from 1 to 2,} %only increase by double,
an extremely large on-off ratio up to $3.72\times10^8\%$ arises.
%For Gr/YIG barriers, the value changes from $3.92\times10^5\%$ at $l_{p,f}=1$
%to $3.72\times10^8\%$ at $l_{p,f}=2$.
%On the other hand, the (spin) Dirac gaps increase for fewer layer magnetic insulators \cite{hallal2017tailoring};
%as a result, an even larger CMR is predicted in graphene under thinner insulators.

%gate-induced insulating state

The strong enhancement can be understood as following.
The insulator state of the device stems from evanescent transports of both spins,
%For evanescent transports the longitudinal wavevector becomes imaginary,
for which the longitudinal wave $e^{ik_mx}$
in Eq.(\ref{wave}) becomes an evanescent wave
$e^{-\kappa x}${,} where $\kappa^2=-k_m^2$.
As a result, %as the ferromagnet length $l$ increases, %and the factor reads $\sim e^{ik_\downarrow l}$.
%as a result,
the transmission and conductance decreases near-exponentially with an increasing barrier length.
Hence, the resistance increases near-exponentially with an increasing length,
see the red solid line in Fig. \ref{fig:length}(b).
%As can be seen, all $R_I$'s increase near-exponentially with increasing
%electrode length, %as a result of the evanescent transport.
In contrast, the metal state stems from a ballistic transport of spin up,
for which $e^{ik_mx}$ is a plane wave. %$e^{ik_mx}$
Accordingly, the transmission, conductance, and resistance
show a weak oscillating dependence on the ferromagnet length,
see the red dashed line in Fig. \ref{fig:length}(b).
% in as a result of ballistic transports.
%and the change in resistance
The enhancement of the on-off ratio follows that of the insulator state.

The cases for devices using EuO and EuS are also plotted in Fig. \ref{fig:length}.
It is seen that, both the {metal-insulator states and the on-off ratio} show similar length dependence
as the device using YIG.
However, the slopes of the profiles are different.
For the device using YIG, the change in resistance increases from $3.92\times10^5\%$ at $l_{p,f}=1$
to $3.72\times10^8\%$ at $l_{p,f}=2$;
for the device using EuO, the {on-off} ratio changes from $3.50\times10^4\%$ to $5.00\times10^6\%$,
while for the device using EuS, the values read $4.08\times10^5\%$ and $7.14\times10^9\%$.
An enhancement factor, $\Delta(\log \textmd{RC})/\Delta l_{f(p)}$,
can be calculated as $\sim$ 3, 2, and 4, respectively.
Interestingly, they are found to be proportional to {the squares} of the smaller effective spin Dirac gaps
(but not the smaller spin Dirac gap itself), i.e., $\Delta \textmd{RC}^{(i)}\propto e^{-(\Delta_\downarrow^{(i)}/v_\downarrow^{(i)})^2\Delta l}$.
This is because the resistance is dominated by the spin with
a smaller $\kappa$ ({i.e.,} spin down for all the three magnetic insulators,
see table \ref{parameter}), %effective Dirac gap, %the factor stems from that
and $\kappa^2\propto
%(\Delta_\downarrow/v_\downarrow)^2$
%$k_m^2+q_m^2=
-\Delta_\downarrow^2/v_\downarrow^2$ near the spin Dirac points.
%which is found to ... with effective gap
%This is because $\Phi(l)$ in all regions ... in Eq. 3 $\sim e^{ik_\downarrow l}$,

\begin{figure}[!t]
  %\flushleft
  \centering
  \includegraphics[width=\linewidth]{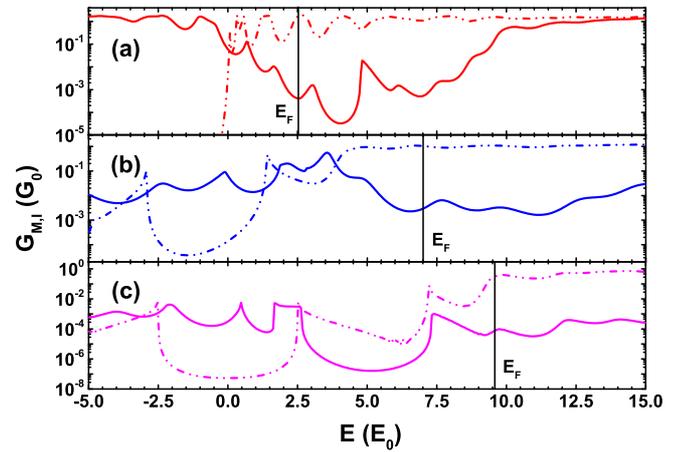}
\caption{%    Metal-like low resistance (hollow), insulator-like high resistance (solid), and CMR
%    as a function of (a) temperature, (c) electrode length, and (d) contact doping
%    for spin valves based on YIG (red), EuO (blue), and EuS (magenta). %, the cases in Fig. \ref{gates} (a-c).
    {The zero-temperature} $G_M$ (dashed) and $G_I$ (solid) as a function of energy
    for devices of $l=1$ based on YIG (a), EuO (b), and EuS (c).
    %, the cases in Fig. \ref{gates} (a-c).
    }\label{fig:conductance}
\end{figure}

\subsection{Temperature dependence of the changes in resistance}

Till now, we only consider the {zero-temperature} on-off ratios.
How it will change at high {temperatures}, especially at {the liquid helium and the liquid nitrogen} temperatures?
In Fig. \ref{Fig:temperature}(b), we plot the numerical results
for $l=2$ (the red solid line) as a function of temperature.
%which is calculated by weight of zero temperature spin conductance.
%Fig. \ref{Fig:temperature}(a) for $l=1$ case we can see that,
%the low resistance increases with increasing temperature while
%the high resistance decreases with increasing temperature;
%the change in resistance decreases sharply.
%they are an important evidence for a metal or insulator behavior.
%these numerical results are calculated by
%the zero-temperature spin-dependent conductance at different configuration
%as a function of energy around the Fermi energy, and make a temperature-dependent weigth ... , see Eq. .
%these behavior can be explained as follows.
%as a function of temperature is calculated through Eq. \ref{},
%by changing the temperature while fixing ... .
%As can be seen in Fig. \ref{temperature}(b),
%%%
As can be seen, the on-off ratio shows a decreasing dependence %high and low resistances
on the temperature, %see Fig. \ref{Fig:temperature}(b). %on the temperature (see, Fig. \ref{temperature}(a2).
which is similar to the temperature dependence {found} in  manganite perovskites
and other Dirac materials.
However, %the CMRs at the liquid helium temperature (4.2K),%%%%
%Due to such a strong enhancement, %for EuO and EuS
the on-off ratio %only decreases by ...percent
maintains $305\times10^6\%$ (18$\%$ smaller than the {value at zero temperature}) at the liquid helium temperature.
This value is hundreds of times higher
than the XMR previously reported %experimental values
in other Dirac materials at several {K} and under magnetic fields of several {T} \cite{ali2014Large,tafti2015resistivity,wang2016resistivity,gao2017extremely}.
The change in resistance maintains as $16\times10^3\%$ at the liquid nitrogen temperature,
which is still  comparable with the CMR observed in manganite perovskites
at the same temperature and under magnetic fields of several {T} \cite{jin1994thousandfold}.
%hundreds of times higher than previously reported experimental values
%%in any other Dirac materials \cite{ali2014Large,wang2016resistivity,tafti2015resistivity,gao2017extremely}
%in conventional materials
%%%under magnetic fields
%\cite{parkin2004giant,yuasa2004giant,chopra2005quantum}
%under magnetic fields of several {T}s.
The Curie temperature of {the} YIG-induce graphene ferromagnet is higher than {the} room temperature.
We have also calculated the on-off ratio at the room temperature by ignoring the inelastic scattering.
A change in resistance of 270$\%$ is found.
%%%%
The temperature dependence for a device of $l=1$ is also shown,
see {the} red dashed line in Fig. \ref{Fig:temperature}(b).
It is found that, the on-off ratios are much smaller and the temperature dependence is much {gentler}.
Moreover, the higher the temperature, the smaller the difference of the {on-off ratios}
for different ferromagnet lengthes.
It is noted that{,} the magnetic-field-induced CMR in manganite perovskites exists only near the
zero-field transition temperature; the above results show that
the proposed electric-field-induced extremely large on-off ratios can survive for a wide range of high temperature.

The {decreasing} temperature behavior of the on-off ratio stems from
the increase behavior of the low-resistance state and the decrease behavior
of the high-resistance state, which are also {important evidences} for a metal or insulator behavior.
In following we will show that
these behaviors stem from a negative and positive
%positive (negative)
energy dependence of the {zero-temperature} conductance around the Fermi energy, respectively,
see Fig. \ref{fig:conductance}.
Due to Eq. (\ref{T-current}), a {spin-dependent} current at a finite temperature $T$ is determined by the
zero-temperature currents in an estimated energy range $\sim (E_F-5T,E_F+5T)$.
%the zero temperature metal-insulator conductance are plotted in Fig. \ref{fig:conductance}.
In Fig. \ref{fig:conductance} we plot the zero-temperature metal-insulator conductances ($G_{M,I}$)
as a function of energy around the Fermi energy.
As can be seen, %in Fig. \ref{temperature} (b),
$G_{M(I)}$ reaches almost the maximum (minimum) around the Fermi energy,
%while ... .
which are direct results of the electronic structure
{as shown in} Fig. \ref{fig:dispersions}.
Accordingly, the higher the temperature, i.e., the wider the energy range,
the smaller (bigger) the finite-temperature $G_{M(I)}$ and $R_{I(M)}$.
%Hence the typical temperature behaviors %as a metal or insulator
%result from
%two opposite energy dependence of the zero-temperature conductance
%around the Fermi energy, which
%this explains the temperature behavior of the ... .
An abnormal decreasing of $R_M$ above $T>58$K is {also}
observed. It stems from a ``W''-shaped $G_M$-$E$ profile. %around the Fermi energy

\begin{figure}[!t]
  %\flushleft
  \centering
  \includegraphics[width=\linewidth]{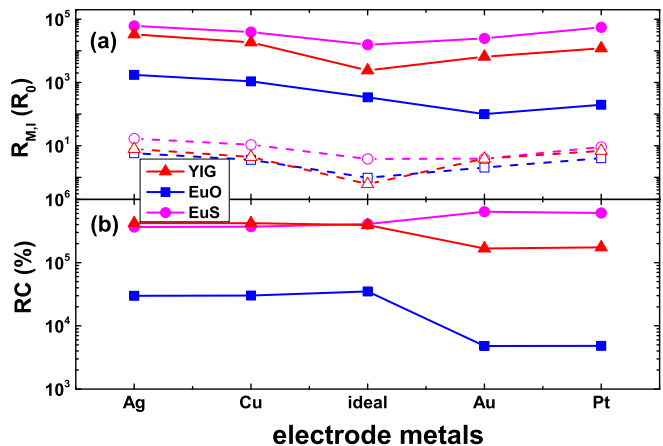}
\caption{ (a) {The} on-state and off-state resistance{s} and (b) {the} change in resistance for
devices based on YIG (red), EuO (blue), and EuS (magenta)
for several contact doping.
    %(b) Zero temperature $G_M$ (dashed) and $G_I$ (solid) as a function of energy
    %for spin valves based on YIG (b1), EuO (b2), and EuS (b3). %, the cases in Fig. \ref{gates} (a-c).
    }\label{fig:contact}
\end{figure}

The devices using EuO and EuS show similar {temperature dependence} (Fig. \ref{Fig:temperature}(b))
and {underlying} mechanisms (Fig. \ref{fig:conductance}).
However, the temperature ranges are much smaller.
At the liquid helium temperature, extremely large changes in resistance up to
%$3.0\times10^8\%$,
$4.9\times10^6\%$ and $6.5\times10^9\%$ are obtained,
%for graphene under YIG, EuO, and EuS,
respectively.

\subsection{Metal contacting effects}

%we adopt a two-probe technique in the proposed device, for which
%the contact resistance cannot be ignored.
At last, we consider the influence of metal contacts on the changes in resistance.
%in Fig. \ref{Fig:contact} we consider %how the CMR will be influenced.
Several familiar metals, Ag, Cu, Au, and Pt at their equilibrium distances with graphene
are considered.
%The calculations are done by change ... ,
%the metal and insulator configuration, spin up and spin down.
%For Ag, Cu, Au, and Pt
The contact {doping's ($U/E_0$) equal}
-32, -17, 19, and 32, respectively \cite{giovannetti2008doping}.
The calculated results are shown in Fig. \ref{fig:contact}.
It is observed that, for device using YIG and EuS,
both $R_I$ and $R_M$ increase as the contacts become non-ideal;
the heavier the contact doping, the larger the resistances increase.
This is because the contacting resistances are in
series with the original ones.
However, the low and high {resistances} increase differently.
As a result, the on-off ratio can show different behaviors.
It decreases for device using YIG (EuS) with positive (negative) doping's, %and for EuO with any doping,
while increases for device using YIG (EuS) with negative (positive) doping's.
Considering that the contact resistance is usually harmful for device performance \cite{russo2010contact}, %nagashio2009metal,
the latter is a rather interesting and useful result.
For device using EuO, the change in resistance always decreases,
slightly for negative contact doping's and sharply for positive contact doping's.

%the EuO and EuS cases ...

\section{Conclusion}

In summary, we have proposed a device that can generate
extremely large changes in resistance at high {temperatures} and in {a}
zero magnetic field.
%through electric-field-induced metal-insulator states.%double barrier
The device is composed of a graphene under two YIG strips, where gate voltages
are applied. %graphene covered by two YIG strips.
{Based on conductance calculations we} have demonstrated that, by applying a proper gate on the free ferromagnet,
an on-off ratio up to $305\times10^6\%$ can be obtained
%(18% smaller than the zero temperature value)
at the liquid helium temperature and in {a} zero magnetic field.
This value is hundreds of times higher than the XMR
%previously reported
{previously} observed in Dirac materials at similar temperatures and under magnetic fields of several {T}.
The change in resistance maintains as $16\times10^3\%$ at the liquid nitrogen temperature and in {a} zero magnetic field,
which is still comparable with the CMR observed in manganite perovskites
at the same temperature and under magnetic fields of several {T}.
%hundreds of times higher than previously reported experimental values in conventional materials.
%ferromagnetic graphene tunneling junction
%resistance difference.
%Compare with the CMR in manganite perovskites and the XMR in Dirac materials,
%the proposed effect needs no magnetic field and survives for high temperature.
We have indicated that, the underlying mechanism for such a remarkable effect is that,
an electric field induces
%... reversal in the electrode
%and evansent transport of both spins in the device.
%an electric-field-induced
%reversal of the full-polarization in the free ferromagnet
a reversal of the full polarization in the half-metallic free ferromagnet,
%or an energy gap in a normal ferromagnet,
which results in metal-insulator states in the device.
%supporting by a ballistic transport of spin up
%and an evanescent transport of both spins.
%an electric-field-induced
%full-polarization reversal %(for graphene under YIG or EuO)
%or insulating state %(for graphene under EuS or EuO)
%in the free electrode,
%which is accompanied by a change from
%a ballistic transport of at least one spin to
%an evanescent transport of both spins.
%The metal-insulator states are found to robust to electrode contacts.

%%%%
Interesting results also contain:
1) the change in resistance shows a near-exponential dependence on the ferromagnet lengths,
2) the longer the {two} ferromagnets, the sharper the negative
{temperature-dependence} of the on-off ratio,
and 3) the effective spin Dirac gap instead of the spin Dirac gap itself plays an important role in the insulator state.
We have also shown that, the proposed effect can be realized
in devices using other magnetic insulators such as EuO and EuS.
%This fact implies that, the electric-field-induced metal-insulator states
%stems from the magnetic proximity effect and electric field effect,
%hence may also be explored in other systems, such as
%graphene on La$_{0.7}$Sr$_{0.3}$MnO$_3$ \cite{sakai2016proximity} and Ni \cite{zhang2017realization};
%monolayer transition metal dichalcogenides %such as
%MoS$_2$ on %cobalt ferrite
%CoFe$_2$O$_4$ (CFO) \cite{jie2017observation} and
%%and on EuS \cite{liang2017magnetic},
%WSe$_2$ on CrI$_3$ \cite{zhong2017van};
%and topological insulators %such as
%Bi$_{2-x}$Mn$_x$Te$_3$ on Fe \cite{vobornik2011magnetic}
%and Bi$_2$Se$_3$ on EuS \cite{katmis2016high}.
%Very recently, intrinsic ferromagnetism %and EFE
%is even found
%in monolayer van der Waals crystals Cr$_2$Ge$_2$Te$_6$ \cite{gong2017discovery,xing2017electric}
%and CrI$_3$ \cite{huang2017layer}.
Our work should be helpful for developing a realistic switching device.
Using an electric field instead of a magnetic field, the proposed device is also
far more energy saving and compatible with the ubiquitous
voltage-controlled semiconductor technology \cite{chiba2008magnetization,ohno2010window,matsukura2015control}.

\section*{Acknowledgements}
I'd like to thank M.S. XL Feng and Dr. SY Hou for inspiring discussions.
This work was supported by the National Natural Science Foundation of China (Grant No. 11404300)
and the Science Challenge Project (Grant No. TZ2016003-1).
%and the IEE S$\&$T Innovation Fund (Grant No. S20140807).

%\bibliographystyle{apsrev4-1} % Tell bibtex which bibliography style to use
%\bibliography{GiCMR}

%merlin.mbs apsrev4-1.bst 2010-07-25 4.21a (PWD, AO, DPC) hacked
%Control: key (0)
%Control: author (72) initials jnrlst
%Control: editor formatted (1) identically to author
%Control: production of article title (-1) disabled
%Control: page (0) single
%Control: year (1) truncated
%Control: production of eprint (0) enabled
%

\end{document}